# A revised uncertainty budget for measuring the Boltzmann constant using the Doppler Broadening Technique on ammonia

C Lemarchand[1,2,*], S Mejri[1,2], P L T Sow[2,1], M Triki[1,2], S K Tokunaga[1,2], S Briaudeau[3], C Chardonnet[2,1], B Darquié[2,1] and C Daussy[1,2,†]

[1] *Université Paris 13, Sorbonne Paris Cité, Laboratoire de Physique des Lasers, F-93430 Villetaneuse, France*

[2] *CNRS, UMR 7538, LPL, F-93430 Villetaneuse, France*

[3] *Laboratoire Commun de Métrologie LNE-CNAM, F-93210, La Plaine Saint-Denis, France*

E-mail: christophe.daussy@univ-paris13.fr

**Abstract:**

We report on our on-going effort to measure the Boltzmann constant, $k_B$, using the Doppler Broadening Technique. The main systematic effects affecting the measurement are discussed. A revised error budget is presented in which the global uncertainty on systematic effects is reduced to 2.3 ppm. This corresponds to a reduction of more than one order of magnitude compared to our previous Boltzmann constant measurement. Means to reach a determination of $k_B$ at the part per million accuracy level are outlined.

---

[*] Present address: Université de Toulouse, UPS, IRSAMC, Laboratoire Collisions Agrégats Réactivité, CNRS UMR 5589, F-31062 Toulouse, France
[†] Author, to whom any correspondence should be addressed.

## 1. Introduction

The International System of units is expected to be redefined at a future meeting of the Conférence Générale des Poids et Mesures (CGPM) [1-6]. At this occasion, a new definition of the kelvin will be implemented by fixing the Boltzmann constant, $k_B$, to the value recommended by the Committee on Data for Sciences and Technology (CODATA). Their 2010 recommendation for the Boltzmann constant is $k_{B(2010)} = 1.380\ 648\ 8(13) \times 10^{-23}$ J.K$^{-1}$ with an uncertainty of 0.9 ppm [7]. This value is based on measurements of the gas constant $R$ and the Boltzmann constant $k_B$ obtained before 2010. Six of the eight measurements reported in ref [7] were published between 2006 and 2010, illustrating the efforts made by several metrology institutes in preparation for the upcoming redefinition of the kelvin. However, most of these results were obtained using a single method, namely Acoustic Gas Thermometry. Our goal is to provide a competitive measurement using the alternative Doppler Broadening Technique (DBT).

In 2002, Ch. J. Bordé proposed the DBT as a new approach for measuring $k_B$ [8]. The principle is to record, by linear spectroscopy, the Doppler profile of an absorption line for a gas in thermal equilibrium. In the Doppler regime, when all other broadening mechanisms are negligible, the recorded Gaussian profile results from the Maxwell–Boltzmann distribution of velocities along the laser beam axis. The e-fold half-Doppler width $\Delta \nu_D$ is directly linked to $k_B/h$ by:

$$\frac{k_B}{h} = \left(\frac{\Delta \nu_D}{\nu_0}\right)^2 \frac{mc^2/h}{2T}, \qquad (1)$$

where $\nu_0$ is the central frequency of the molecular line, $c$ is the speed of light, $h$ the Planck constant, $T$ is the temperature of the gas and $m$ its molecular mass. The relative uncertainties on $mc^2/h$ and $h$ are $10^{-8}$ (deduced from atom interferometry experiments with rubidium [9] and atomic mass ratios measured in ion traps [10]) and $5\times10^{-8}$ (deduced from the Watt balance experiment [7]), respectively. We measured $\nu_0$ with a relative uncertainty of a few $10^{-9}$ using saturated absorption spectroscopy [11]. The uncertainty on $k_B$ is then dominated by uncertainties on the temperature and the Doppler width. In the DBT, the temperature and line shape are thus accurately measured simultaneously. The Doppler width is extracted from the line shape using numerical nonlinear regression. $k_B$ is then calculated from Equation (1).

At the Laboratoire de Physique des Lasers (LPL), we have followed this new approach to measure $k_B$ by carrying out laser spectroscopy of an ammonia rovibrationnal line at 10.35 μm. In 2007, a first measurement was completed with an uncertainty of 190 ppm [12-14]. Four other groups then started to develop similar experiments using $CO_2$, $H_2O$, $C_2H_2$ and Rb [15-20]. By investigating and improving the control over some of the systematic effects, we subsequently made an improved measurement with an uncertainty of 50 ppm. Presently the most accurate value for the Boltzmann constant that can be deduced from Doppler spectroscopy is $k_B= 1.380704(69) \times 10^{-23} J.K^{-1}$ [11]. This result was reported in the 2010 *CODATA Recommended Values of the Fundamental Physical Constants* [7] as a new result "of interest because it is obtained from a relatively new method that could yield a value with a competitive uncertainty in the future". Our 2007 and 2010 uncertainty budgets were dominated by the uncertainty on $\Delta \nu_D$, which was in turn limited by our knowledge of the absorption line shape, an issue also encountered by other groups [19, 21-23].

To reach the ppm level, we have conducted a refined analysis of the main sources of systematic uncertainty. In section 2, the experimental setup is described. In sections 3 and 4, we focus on uncertainties associated with the determination of $T$ and $\Delta \nu_D$, respectively. Finally section 5 concludes the paper with a detailed uncertainty budget showing a reduction of the systematic uncertainty in our measurement of the Boltzmann constant by more than one order of magnitude.

## 2. Experimental setup

The experimental setup is shown in figure 1.

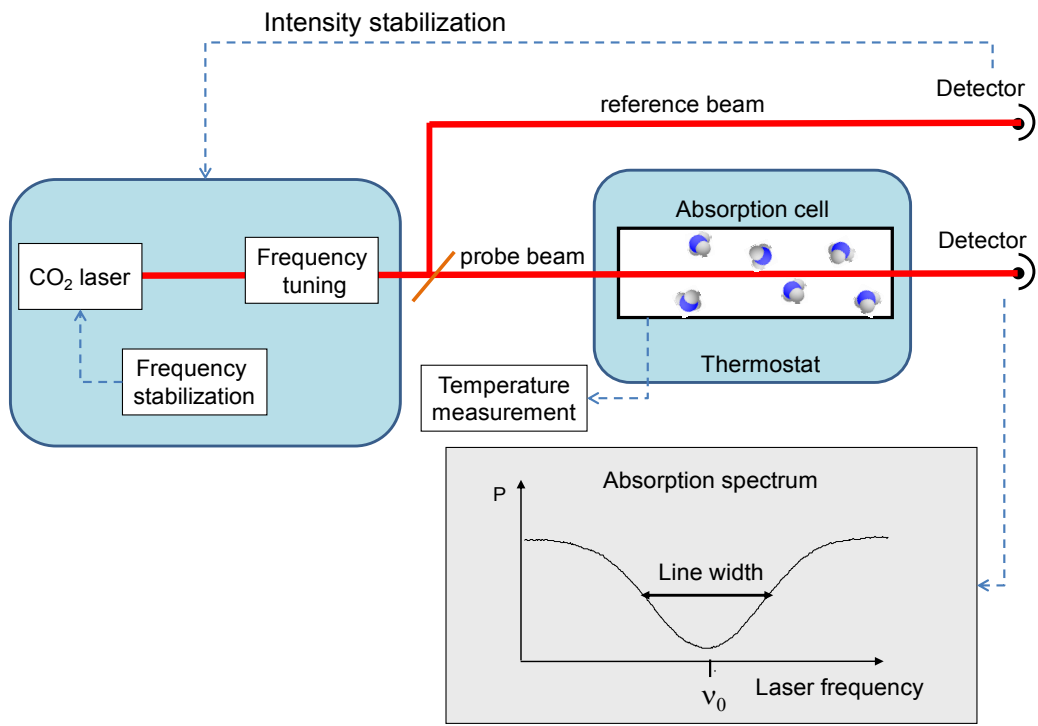

Figure 1: Experimental setup principle for $k_B$ measurement by the DBT.

The laser source, a frequency stabilized $CO_2$ laser around 10 μm, exhibits a spectral width smaller than 10 Hz and a frequency stability of 0.1 Hz ($3 \times 10^{-15}$) for a 100 s integration time [24]. The laser beam is split into two parts: a reference beam and a probe beam. The reference beam is used to stabilize the intensity of the probe beam to ensure that the line shape is measured at constant incident power. The laser frequency is scanned to record the profile. The absorption length of the cell is 37 cm in a single-pass configuration (SPC) or 3.5 m in a multipass configuration (MPC). The linear absorption of the chosen transition, the $\nu_2$ saQ(6,3) line of $^{14}NH_3$ (of central frequency $\nu_0 = 28\ 953\ 693.9(1)$ MHz), varies from 15% to 98% for pressures ranging from 1 to 25 Pa in SPC and from 0.1 to 2.5 Pa in MPC.

This experiment requires the molecular gas to be maintained at a constant and homogeneous temperature. The absorption cell is placed inside a copper thermal shield which is itself inside a stainless steel enclosure immersed in a thermostat filled with an ice-water mixture which stabilizes the temperature close to 273.15 K (see section 3).

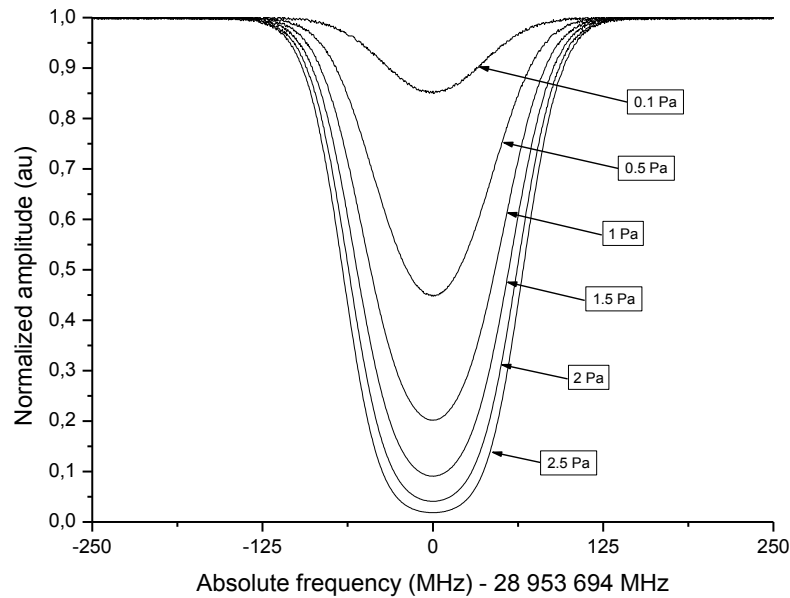

Figure 2: Spectra of the $\nu_2$ saQ(6,3) line of $NH_3$ recorded in the MPC for pressures ranging from 0.1 to 2.5 Pa.

Typical spectra recorded at ammonia pressures ranging from 0.1 to 2.5 Pa in the MPC, are displayed in figure 2. They are about 100 MHz wide, recorded over 500 MHz by steps of 500 kHz with a 30-ms integration time per point, leading to a typical signal-to-noise ratio of $10^3$.

## 3. Temperature control

The gas temperature is measured with three small (5 cm-long) 25 Ω Hart capsule Standard Platinum Resistance Thermometers (cSPRTs) in thermal contact with the absorption cell. These thermometers are calibrated at the triple point of water and at the gallium melting point at the Laboratoire Commun de Métrologie LNE-CNAM. In order to measure the cell temperature, they are monitored with a resistance measuring bridge continuously calibrated against a resistance standard with a very low-temperature dependence. As the bridge and the standard resistance used for cSPRTs calibration are different from those used to measure the temperature of the gas cell, one has to take into account their reproducibility. The combined uncertainty on the temperature measurement has various origin that have been investigated: the standard resistance calibration ($R_{std}$), the cSPRTs resistance calibration at the water triple point $R(273.16\ K)$, the self-heating correction caused by the 1 mA electrical current applied to cSPRT's during measurements ($C(T)$) and the bridge resistance ratio measurement accuracy at 1mA ($ratio(T, 1\ mA)$). Thermal probes are very fragile and their stability over one year is about 0.8 mK. If calibrated just before and after 3 months of use in the DBT experiment, their calibration uncertainty can be reduced down to 0.2 mK.

Within 0.1 K off the water triple point temperature, one can approximate the temperature measured with cSPRTs from the following expression [25]:

$$T\,(K) = 273.16 + s\left[\frac{ratio(T,1\,mA)*R_{std}+C(T)}{R(273.16\,K)} - 1\right] \qquad (2)$$

Where $s$, the "standard" temperature sensitivity of the platinum resistance close to the water triple point temperature is equal to $250.7190\,K$ as given by the International Temperature Scale 1990 (ITS90). $C(T)$ is the self-heating correction defined as $R(T,0mA) - R(T,1mA)$. In these conditions, the uncertainty on the measured temperature with our set-up is approximated by [25]:

$$\sigma(T) \sim s\sqrt{\left[\frac{\sigma(ratio(T,1\,mA))}{ratio(T,1mA)}\right]^2 + \left[\frac{\sigma(R_{std})}{R_{std}}\right]^2 + \left[\frac{\sigma(R(273.16\,K))}{R(273.16\,K)}\right]^2 + \left[\frac{\sigma(C(T))}{R(273.16\,K)}\right]^2} \qquad (3)$$

From Table 1, the combined uncertainty on the cSPRT temperature measurement is estimated to be 0.3 mK.

| Component | Value | uncertainty (k=1) | Sensitivity | Temperature uncertainty (k=1) | Comment |
|---|---|---|---|---|---|
| $R_{std}$ | 10.000516 Ω | 8 μΩ | 25 K.$\Omega^{-1}$ | 0.2 mK | Reproducibility of standard resistance |
| R (273.16 K) | 25.517610 Ω | 20 μΩ | 9.8 K.$\Omega^{-1}$ | 0.2 mK | cSPRT calibration at water triple point (0 mA) |
| ratio (T, 1mA) | 2.5519369 | $1.0\times10^{-6}$ | 98 K | 0.1 mK | Reproducibility of bridge ratio (1 mA) |
| C(T) | -68 μΩ | 5 μΩ | 9.8 K.$\Omega^{-1}$ | 0.05 mK | Self-heating correction measured on the absorption cell |
| **T** | **273.18955 K** | | | **0.3 mK** | **Combined standard uncertainty** |

Table 1: Uncertainty budget (k=1) of the cSPRTs temperature measurement.

The thermal shield (see section 2) is linked to its enclosure via only a copper ring. The absorption cell is mechanically linked to the thermal shield via Teflon holders to ensure that heat transfer between them is predominantly radiative (the whole system is under vacuum). This in turn ensures that the heat transfer is spatially and temporally (~7 h thermal time constant) smoothed, thereby reducing residual temperature gradients and drifts. Temperature drifts in the cell have been measured to be less than 273 μK per day, making them completely negligible during a single laser scan, which typically lasts 1 min. Typical noise on the temperature measurement around 273.15 K (corresponding to the temperature statistical uncertainty) is displayed in figure 3, after correction for a linear drift. For 1 min integration time, the detected noise is less than 6μK.

Our purpose-specific thermostat requires optical windows. Blackbody radiation from the laboratory at room temperature (300 K) heats the gas cell walls and induce a temperature gradient over the cell of about 1 mK. The cell temperature inhomogeneity has been investigated by placing the three probes at various positions on the faces of the cell. Using interference filters and optical diaphragms, the temperature inhomogeneity is reduced. Measurements show a reproducible horizontal gradient of 11 μK.cm$^{-1}$, with a corresponding uncertainty (k=1) on cell temperature of 229 μK (within the temperature measurement uncertainty of 0.3 mK) while the vertical temperature inhomogeneity, over the laser beam diameter, is negligible in comparison. More details on the thermostat and on the temperature control can be found in [25].

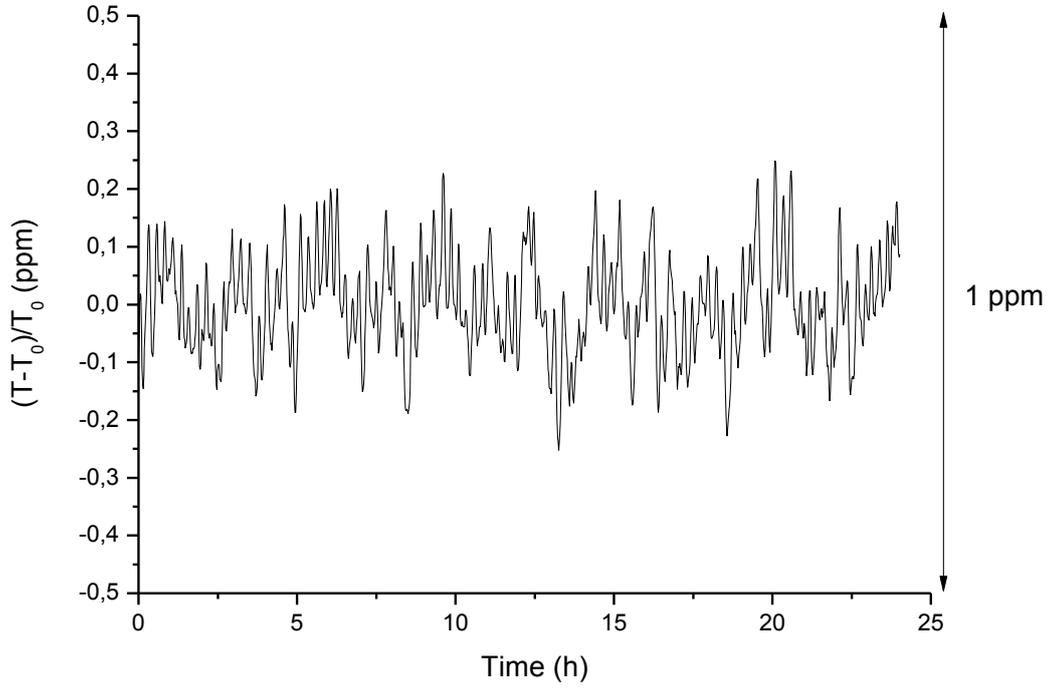

Figure 3: Absorption cell temperature measurement noise around $T_0$=273.15 K (a linear temperature drift of -8.7 µK/h has been corrected for). Data recorded over 24 h with a 25 Ω capsule Standard Platinum Resistance Thermometer (cSPRT) with an integration time of 60 s.

To conclude, temperature stability and inhomogeneity, cSPRTs' calibration and calibration of the temperature measurement add up to an uncertainty on the probed gas temperature of 0.38 mK. The statistical uncertainty on temperature measurement remains negligible.

## 4. Doppler width measurement

In this section, we discuss the uncertainties associated to a number of systematic effects that may affect the determination of the Doppler width, $\Delta\nu_D$. The latter is obtained by fitting recorded spectra to the Beer-Lambert law (with an added baseline):

$$P(\omega) = (P_0 + P_1(\omega - \omega_0)\exp[-A.I(\omega)] , \qquad (4)$$

with parameters $P_0$, $P_1$, $\omega$, $\omega_0$ and $A$ respectively corresponding to the offset and slope of the baseline, to the angular frequency of the laser and of the molecular line center, and to the integrated absorbance, proportional to the partial pressure of ammonia. $I(\omega)$ is the normalized absorption profile which contains the Doppler width, $\Delta\nu_D$.

As already mentioned, until recently, the main source of uncertainty on determining $k_B$ was limited by our knowledge of the absorption line shape and our ability to reliably fit the experimental data [11, 14, 26-28]. For instance, it constituted 97% of the 144 ppm standard uncertainty published in [11].

## 4.1. The different line shape models

When the pressure goes to zero Pa, the collision rate vanishes. The absorption profile $I(\omega)$ (see Eq. 4) is purely Doppler broadened and is a simple Gaussian of e-fold half-width $\Delta\nu_D$, of about 50 MHz in our experimental conditions (the natural width for rovibrational levels is negligible, of the order of a few Hz). In practice however, collisions induce both an additional broadening $\Gamma$ and a shift $\Delta$. The overall line shape is then a Voigt profile (VP), *i.e.* the convolution of the above mentioned Gaussian with a Lorentzian. In our experimental conditions the collisional broadening and shift (the actual half-width at half-maximum and shift of the Lorentzian) are typically 100 kHz/Pa and 1 kHz/Pa respectively. The VP is probably the most widely used in the spectroscopy community. However more subtle effects cannot be ignored in our case. The collisional broadening and shifting actually slightly depend on the molecular speed. Taking this dependence into account leads to the speed-dependent Voigt profile (SDVP) with an overall narrowing of the line shape. Moreover collisions induce a confinement of molecules by the surrounding molecules which may lead to a narrowing of the Doppler contribution to the line shape [29]. This narrowing is important when the pressure is such that this confinement is at the scale of the wavelength, the so-called Lamb-Dicke-Mössbauer (LDM) regime. In our case the LDM narrowing is expected to be about $\beta \sim 10$ kHz/Pa [30], where $\beta$ is the frequency of velocity-changing collisions, an additional parameter of the line shape. In the general case, the line profile can result from the combination of all these effects leading to more complex shapes [31].

### 4.2. Optimal experimental conditions

In principle, it is desirable to work at low pressures to avoid the above mentioned collisional effects which distort the line shape. Our $k_B$ values recently published [11, 26, 27] were thus determined from spectra recorded over 250 MHz at pressures typically below 1 Pa (and never above 2.5 Pa), at which the collisional broadening $\Gamma$ was not negligible, but deviations from the VP were very small. However such experimental conditions did not enable to easily distinguish the homogenous component $\Gamma$, from the Doppler component, $\Delta\nu_D$, during the fit procedure. To make the fitting algorithm converge, the total and partial pressures in the cell were assumed to be equal, allowing us to rewrite $\Gamma$ as $\Gamma_0 \times p$, with $p$ the pressure left as an adjustable parameter and proportional to the integrated absorbance $A$ (see Eq. 4). $\Gamma_0$ was fixed and constrained to a constant value for all fitted spectra. This value was determined following an iterative procedure so that, as expected, the adjusted values of $\Delta\nu_D$ show no dependence with pressure. Both the VP [27] and profiles including LDM narrowing (however fixing $\beta$ to values found in the literature) [11,26] could be tested for fitting to our experimental data. However, the assumption on the gas composition prevented us from taking account of both the residual base pressure and outgassing (during spectra acquisition) giving rise to a possible systematic shift on $\Delta\nu_D$ (as mentioned in [32]). In order to get some insight in these pressure related effects, spectra recorded at low pressure (between 0.15 Pa and 0.85 Pa in MPC) were fitted to VP fixing $\Delta\nu_D$ to the value determined by the temperature of the gas so that $A$ and $\Gamma$ can be left as free parameters. Analysis of the dependence of $\Gamma/A$ versus time and pressure indicated minimal outgassing but a clear signature of a background pressure potentially leading to a 100 ppm systematic shift in the very low covered pressure range [33].

To overcome this limitation, new experimental developments led to a reduced residual base pressure (two orders of magnitude below the ammonia filling pressure) and to the ability to record spectra over 500 MHz. This enabled an accurate fit of the wings of the profile, where the Lorentzian behaviour dominates. Keeping the pressure between 10 and 20 Pa in SPC, and 1 and 2 Pa in MPC (absorption between 80% and 100%) was then found to correspond to optimal experimental conditions under which $\Delta\nu_D$ and $\Gamma$ could be adjusted simultaneously for each individual spectrum. The high-pressure limit is such that total absorption is avoided and signal-to-noise ratio considerations set the low-pressure limit. Under such ideal experimental conditions, systematic effects on the Doppler width due to gas composition can be avoided at the ppm level.

### 4.3. Molecular collisions

Knowing the correct collisional model for the gas, and therefore the true line shape of a given absorption line is difficult. In the optimal experimental conditions detailed in the previous paragraph, deviations from the VP cannot be neglected. We have recently carried out an extensive line shape study at relatively high pressure (from 2 to 20 Pa), in order to enhance collisional effects and to determine the most suitable model for our data [28]. In the covered pressure range, the deviation of the saQ(6,3) line of $^{14}NH_3$ from a VP is dominated by speed-dependent effects and not by LDM effects. Velocity-changing collisions are thus ignored leading to a normalized absorption profile of the form:

$$I_{\text{SDVP}}(\omega) = \frac{1}{\pi} Re \left( \int \frac{F_{\text{M}}(\vec{v})}{\{\Gamma(v) - i[\omega - \omega_0 - \Delta(v) - \vec{k}.\vec{v}]\}} d^3\vec{v} \right) \quad (5)$$

Where $F_{\text{M}}(\vec{v})$ is the Maxwell-Boltzmann distribution of velocities, $\vec{k}$ the wave vector, and $\vec{v}$ and $v$ are respectively the molecular velocity and its modulus. $\Delta(v)$ and $\Gamma(v)$ are the collisional shift and broadening that depend on $v$.

Residuals obtained for spectra fitted to a SDVP and recorded around 17 Pa are displayed in figure 4 showing a good agreement between this model and experimental line shapes. Residuals obtained after a fit to a VP are also shown for comparison.

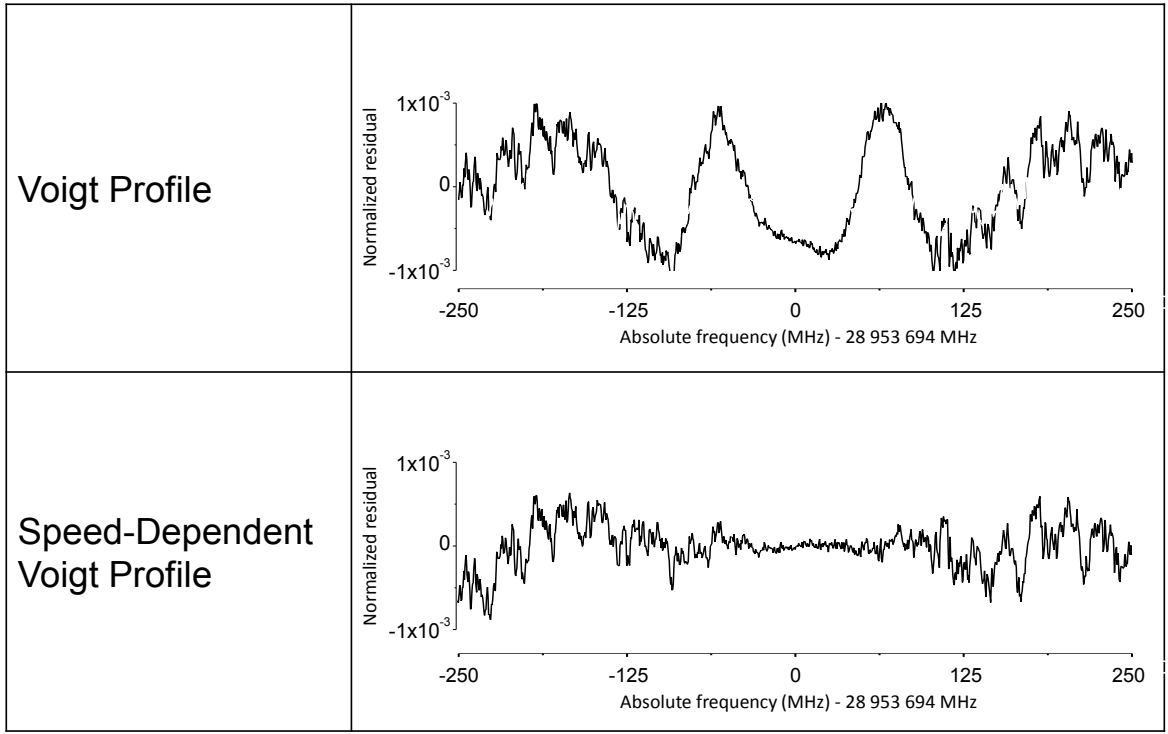

Figure 4: Normalized residuals for non-linear least-squares fits of spectra recorded at 17.3 Pa, to a VP and a SDVP. The frequency scale is offset by 28 953 694 MHz.

The following speed-dependent line shape parameters have been obtained at the few percent level: $\mathrm{d}\Gamma(v)/\mathrm{d}p = 120(3)$ kHz/Pa, $\mathrm{d}\Delta(v)/\mathrm{d}p = 1.2(1)$ kHz/Pa, $m_{SD} = 0.360(7)$ and $n_{SD} = $ - 3.8(3). The $m_{SD}$ and $n_{SD}$ parameters are related to the speed-dependence of the collisional broadening and shift [28].

Having extracted the line shape parameters $m_{SD}$ and $n_{SD}$ with good accuracy from data taken at high pressure, we fix them when fitting the low pressure data used to determine $k_B$. The uncertainty on the Doppler width due to collisional effects is then only limited by our knowledge of the associated parameters $m_{SD}$ and $n_{SD}$. Simulated SDVP spectra corresponding to the optimal experimental conditions in MPC (pressure below 2 Pa) have thus been fitted taking into account uncertainties on $m_{SD}$ and $n_{SD}$. We followed the line-absorbance based analysis method recently proposed in [34]. By applying this method on the simulated spectra, we obtained an expected experimental uncertainty of 0.9 ppm on the Doppler width measurement [28]. In order to quantitatively estimate the impact of the line shape model on the Doppler width measurement accuracy, the previous simulated SDVP spectra have also been fitted with a VP. In this case, an expected underestimation of the Doppler width leads to a systematic effect of -8.6 ppm. The expected experimental uncertainty of 0.9 ppm obtained with a SDVP represents a reduction by one order of magnitude, of the main source of uncertainty in the measurement of the Boltzmann constant with our experiment.

### 4.4. Transit-time

It is sometimes believed that there should be an additional broadening of the line shape owing to the finite transit time of particles through the laser beam. It has recently been shown that this is not the case in linear absorption spectroscopy, provided that the medium is uniform and isotropic [35]. Furthermore, this absence of transit time broadening holds true even after introducing a speed-dependent broadening and shift in the calculations [36]. We outline here the derivation given in [35, 36]. The starting point is the expression for the dimensionless absorbance in the momentum representation, $A(\omega) = A.I(\omega)$:

$$A(\omega) \propto \frac{L}{V} \int d^3k \, \alpha^*(\vec{k}) \, \alpha(\vec{k}) \, \bar{f}(\vec{k}, \omega) + \text{c. c.} \tag{6}$$

with $L$ the absorption length and $V$ the laser mode volume. The normalized line shape for each plane wave component $\alpha(\vec{k})$ of the laser mode is the velocity average $\bar{f}(\vec{k}, \omega) = \int f(\vec{v}, \vec{k}, \omega) d^3v$ where $f(\vec{v}, \vec{k}, \omega)$ is proportional to the first-order off-diagonal density matrix element $\rho_{ab}^{(1)} = i\Omega_0 f$ through the Rabi frequency $\Omega_0$ and satisfies the integral equation:

$$\left[ i(\omega - \omega_0 - \Delta(v) - \vec{k}.\vec{v}) + \Gamma(v) \right] f(\vec{v}, \vec{k}, \omega) = F_\mathrm{M}(\vec{v}) + \int d^3v' \, W_0(\vec{v'} \rightarrow \vec{v}) \, f(\vec{v'}, \vec{k}, \omega) \tag{7}$$

Where $W_0(\vec{v'} \rightarrow \vec{v})$ is the collision kernel describing the probability for a velocity change from $\vec{v'}$ to $\vec{v}$, which can potentially give rise to LDM narrowing.

The wave vector only appears in this equation through the projection $\vec{k}.\vec{v}$ such that, thanks to the isotropy of the velocity distribution and of collisions, there is no privileged direction. Thus $\bar{f}$ cannot depend on the direction of $\vec{k}$ but only on its modulus $|\vec{k}|$. Owing to the dispersion relation $k = \omega/c$, $\bar{f}$ is therefore only a function of $\omega$. The mode volume $V$ cancels with $\int d^3k \, \alpha^*(\vec{k}) \, \alpha(\vec{k})$ and any dependence on the laser mode content disappears. The resulting expression for the absorbance is independent of the light beam geometry, and hence has no transit-time broadening.

### 4.5. Modulation broadening

For noise filtering, both the reference and the probe beam are amplitude modulated at $f_1 = 40$ kHz (modulation index of 1) and the signal is detected after demodulation at $f_1$. The laser field probing the ammonia gas is thus composed of 3 components at angular frequencies $\omega$, $\omega + f_1$ and $\omega - f_1$. The presence of these sidebands in the laser spectrum leads to a broadening of the line shape which has been evaluated by simulation of the first harmonic of the resulting signal detected at $f_1$. The simulated line shape was then fitted by a unique SDVP (as is done when analysing data). The laser beam modulation broadening leads to a pressure-independent overestimation of the Doppler width of +0.23(2) ppm which can be taken into account and compensated for.

*4.6. Hyperfine structure*

The saQ(6,3) line has been chosen because it is a well-isolated rovibrational line. However, owing to the non-zero spin values of the N and H nuclei, an unresolved hyperfine structure is present in the Doppler profile causing the overall line shape to deform and broaden according to the relative position and strength of the 78 individual hyperfine components. An accurate knowledge of this structure is required in order to apply the right correction to the Doppler width.

A combined analysis of infrared saturated absorption spectra of the saQ(6,3) [11] and microwave data from the literature [37-39] led to an accurate determination of the hyperfine constants of the transition. The position of the 78 hyperfine components and their relative intensities were then calculated, together with the associated uncertainties. In order to estimate this hyperfine structure-induced systematic effect, we fit to a single SDVP a simulated spectrum composed of the sum of 78 SDVP with positions and intensities corresponding to the precisely determined hyperfine structure. This leads to an overestimation of $\Delta v_\mathrm{D}$ of +4.356 (13) ppm which can thus be taken into account.

To conclude this section, no systematic effect due to the laser beam geometry is expected on the Doppler width measurement. The uncertainty coming from collisional effects is limited by our knowledge of the spectroscopic parameters, at 0.9 ppm. Finally, the laser beam modulation and the unresolved hyperfine structure are responsible for an overestimation of the Doppler width that can be corrected for at the level of respectively 0.02 and 0.013 ppm. Since the Doppler width depends on the square root of $k_B$, the corresponding uncertainties on the Boltzmann constant are twice larger.

## 5. Summary and prospects

Considering the various contributions discussed in section 3 and 4, an uncertainty budget on the systematic effects affecting the Boltzmann constant measurement can be drawn.

| Component | Relative uncertainty on $k_B$ (parts in $10^6$) | Comment |
|---|---|---|
| **Doppler width** | | |
| | | |
| Type B estimates: | | |
| Natural width | Negligeable | |
| Gas composition | Negligeable | |
| Laser beam geometry | No impact | |
| Absorption line shape | 1.8 | Uncertainty on $m_{SD}$ and $n_{SD}$ |
| Laser beam modulation | 0.04 | Numerical simulations |
| Hyperfine Structure | 0.03 | |
| Laser linewidth | Negligeable | Deduced from the laser spectral width |
| Linearity and accuracy of the laser frequency scale | Negligeable | Deduced from the frequency tuning technique |
| | | |
| **Temperature** | | |
| | | |
| Type B estimates: | | |
| Drift | 0.0007 | During 1 spectrum acquisition (1 min) |
| Inhomogeneity | 0.84 | Absorption cell inhomogeneity |
| $T$ measurement | 1.1 | LCM-LNE-CNAM estimation |
| | | |
| **Combined standard uncertainty** | **2.3** | **Root sum of squares** |

Table 2 : Uncertainty budget (k=1) for the determination of the Boltzmann constant by the DBT at LPL for spectra recorded between 1 and 2 Pa in MPC.

From table 2 it can be concluded that a measurement of $k_B$ by the Doppler Broadening Technique at a 2.3 ppm accuracy is now reachable. This corresponds to an improvement of more than one order of magnitude in the systematic uncertainty of our measurement of the Boltzmann constant.

This overall uncertainty budget does not take into account any possible nonlinearity of the detection setup and possible saturation of the probed transition, as no effect has been observed at the current experimental sensitivity. Nevertheless, a new approach, based on a tunable thermostat, is planned in order to cancel this eventual source of systematic effect. The main advantage of the current melting ice thermostat is its simplicity and its intrinsic temperature stability. However, it is not possible to change the temperature of the bath. To enable a variable but stable working temperature, the melting ice will be replaced by a 1 m³ mixture of water and alcohol, maintained at a desired temperature. A cryostat actively coupled to a heat exchanger will allow a regulation of the temperature of the thermostat liquid bath from +10 °C down to −10 °C. In this temperature range, the relative uncertainty associated with the interpolated temperature of the used cSPRTs remains within 1 ppm. Using this new setup, the temperature dependence of the Doppler width will be measured in order to deduce the Boltzmann constant from the slope of $\Delta\nu_D(T)$. For spectra recorded at a given pressure and laser intensity, any nonlinearity of the detection chain would be (independent of $T$) observed as an additional offset on $\Delta\nu_D(T)$.

To improve the line shape profile analysis, a complementary study at higher pressures is under progress and will enable us to refine our understanding of collisional effects. After having extracted spectroscopic parameters with good accuracy from data taken at high pressure, they will be fixed when fitting the low pressure data used to determine $k_B$. Our goal is to reduce to below 1 ppm the uncertainty associated to the line shape knowledge. A new 3-cm long absorption cell has been designed to record spectra at pressures up to 200 Pa. In this pressure range both speed-dependent and LDM mechanisms should significantly contribute to the narrowing of the line shape.

The $CO_2$ laser based spectrometer presents two major limitations, its complexity (see Ref. [11] and [24]) and a limited frequency tunability. In order to overcome these limitations we are currently developing a compact spectrometer based on a widely tunable laser source - a Quantum Cascade Laser (QCL). The tunability of the spectrometer will be increased by more than three orders of magnitude. Together with the higher available intensity and the potentially lower amplitude noise of the QCL source, a large reduction of the time needed to record absorption spectra (limited by the currently used complexity of the $CO_2$ laser frequency tuning technique) is expected. Such improvements will contribute to a reduction of the statistical uncertainty on $\Delta\nu_D$ and, in turn, to a reduction of the time needed to reach the ppm level on $k_B$.

**Acknowledgements**

This work is funded by CNRS, the Laboratoire National de Métrologie et d'Essais and by the European Community (EraNet/IMERA). The authors would like to thank Y. Hermier, F. Sparasci and L. Pitre from Laboratoire Commun de Métrologie LNE-CNAM for platinum resistance thermometers calibrations. We also thank Ch. J. Bordé for many fruitful discussions over the course of this project.